\documentstyle[aps,prb]{revtex}

\newcommand{\supscrpt}[2]{{#1}^{{#2}}}
\newcommand{\subscrpt}[2]{{#1}_{{#2}}}
\newcommand{\mixten}[3]{{#1}^{#2}_{\phantom{{#2}} #3}}

\newcommand{\bfc}{{\bf c}}
\newcommand{\bfcsub}[1]{\subscrpt{\bfc}{#1}}
\newcommand{\bfci}{\bfcsub{i}}

\newcommand{\bfx}{{\bf x}}
\newcommand{\bfxci}{\bfx + \bfci}
\newcommand{\bfxt}{(\bfx, t)}

\newcommand{\bge}{\begin{equation}}
\newcommand{\ee}{\end{equation}}
\newcommand{\bgc}{\begin{center}}
\newcommand{\ec}{\end{center}}
\newcommand{\bgea}{\begin{eqnarray}}
\newcommand{\eea}{\end{eqnarray}}
\newcommand{\bgeas}{\begin{eqnarray*}}
\newcommand{\eeas}{\end{eqnarray*}}

\newcommand{\advop}[2]{\mixten{{\cal A}}{#1}{#2}}

\newcommand{\kop}[2]{\mixten{K}{#1}{#2}}

\newcommand{\vop}[2]{\mixten{V}{#1}{#2}}

\newcommand{\fsup}[1]{\supscrpt{f}{#1}}
\newcommand{\gsup}[1]{\supscrpt{g}{#1}}
\newcommand{\nsup}[1]{\supscrpt{n}{#1}}
\newcommand{\nixt}{\nsup{i}\bfxt}

\newcommand{\nnsup}[1]{\supscrpt{N}{#1}}

\newcommand{\nnixt}{\nnsup{i}\bfxt}

\newcommand{\ggamsup}[1]{\supscrpt{\Gamma}{#1}}

\newcommand{\phisup}[1]{\supscrpt{\Phi}{#1}}

\newcommand{\dt}{\Delta t}
\newcommand{\tpdt}{t+\dt}

\def\glossitem#1!#2!#3{$#1$ \> \if !#2! \else \pageref{#2} \fi\>
    \parbox[t]{3.5truein}{#3} \\}

\def\registeredsymbol{
  {\ooalign{\hfil\raise.05ex\hbox{\hskip.02ex$\rm
   \scriptscriptstyle R$}\hfil\crcr
   \hbox{$\scriptstyle\mathchar"20D$}}}}
\def\registered{
  \ifmmode^\registeredsymbol\else$^\registeredsymbol$\fi}

\def\vertex#1#2#3#4{\begin{picture}(8,8)(- 4,- 4)
\put(0,0){\circle*{3}}
\ifnum#1=1 \put(0,0){\line(1,0){4}}\fi
\ifnum#2=1 \put(0,0){\line(0,1){4}}\fi
\ifnum#3=1 \put(0,0){\line(-1,0){4}}\fi
\ifnum#4=1 \put(0,0){\line(0,-1){4}}\fi
\end{picture}}

\input epsf

\begin{document}

\title{
\begin{flushleft}
{\footnotesize BU-CCS-941102, MIT-CTP-2385}\\
{\footnotesize comp-gas/9501002}\\[0.3in]
\end{flushleft}
Renormalized Equilibria of a Schl\"{o}gl Model Lattice Gas\footnote{
\small \baselineskip=11pt This work was supported in part by the
divisions of Applied Mathematics of the U.S. Department of Energy (DOE)
under contracts DE-FG02-88ER25065 and DE-FG02-88ER25066, and in part by
the U.S. Department of Energy (DOE) under cooperative agreement
DE-FC02-94ER40818.}
}
\author{
Bruce M. Boghosian\\
{\small \sl Center for Computational Science,}\\
{\small \sl Boston University,}\\
{\small \sl 3 Cummington Street, Boston, Massachusetts 02215, U.S.A.} \\
{\small \tt bruceb@conx.bu.edu} \\
[0.3cm]
Washington Taylor\\
{\small \sl Center for Theoretical Physics,}\\
{\small \sl Laboratory for Nuclear
Science and Department of Physics,}\\
{\small \sl Massachusetts Institute of
Technology; Cambridge, Massachusetts 02139, U.S.A.} \\
{\small \tt wati@mit.edu} \\
[0.3cm]
}
\date{December 1, 1994}
\maketitle

\begin{abstract}
A lattice gas model for Schl\"{o}gl's second chemical reaction is
described and analyzed.  Because the lattice gas does not obey a
semi-detailed-balance condition, the equilibria are non-Gibbsian.  In
spite of this, a self-consistent set of equations for the exact
homogeneous equilibria are described, using a generalized
cluster-expansion scheme.  These equations are solved in the
two-particle BBGKY approximation, and the results are compared to
numerical experiment.  It is found that this approximation describes the
equilibria far more accurately than the Boltzmann approximation.  It is
also found, however, that spurious solutions to the equilibrium
equations appear which can only be removed by including effects due to
three-particle correlations.
\end{abstract}

\vspace{0.375truein}



\section{Introduction}

Lattice gas automata have been widely used as models of nonequilibrium
statistical systems since it was shown in 1986 that they could be used
to model Navier-Stokes fluids~\cite{bib:fhp}.  Lattice gases consist of
particles moving about and colliding on a lattice in such a way that
their macroscopic behavior satisfies hydrodynamic partial differential
equations.  Like the Ising model, they are simple discrete systems which
are well suited both to computer implementation and to elegant analytic
techniques; unlike the Ising model, however, they can be used to study
phenomena far from equilibrium.

All of the usual tools of kinetic theory can be used for the analysis of
lattice gases.  Lattice gases whose collisions obey a condition known as
{\it semi-detailed balance} (SDB) can be shown~\cite{bib:fchc} to have a
Gibbsian (product) equilibrium distribution.  As the lattice spacing
goes to zero, expansion about this equilibrium yields the hydrodynamic
equations satisfied by the system; this is a discrete version of the
usual Chapman-Enskog procedure~\cite{bib:swolf}.

To date, most analyses of lattice gases have been done using the
Boltzmann molecular chaos approximation.  We have recently used cluster
expansion methods to develop an exact description of SDB lattice
gases~\cite{bib:long}.  In such lattice gases, the exact equations of
motion include the effects of correlations which renormalize the lattice
gas transport coefficients.  In this paper, we extend these methods to
describe a particular non-semi-detailed-balance (NSDB) lattice gas.
Related work on exact equations for NSDB lattice gases has recently been
done by Bussemaker et. al.~\cite{bib:bed}.

It has been known for decades that chemically reacting systems far from
equilibrium can exhibit fascinating phenomenology, including pattern
formation~\cite{bib:turing} and symmetry breaking~\cite{bib:prigogine}.
Such complicated phenomenology can arise from very simple chemical
reactions, and idealized model reactions have been developed to
illustrate these phenomena.  For example, the simple model reaction
proposed by Schl\"{o}gl in 1972~\cite{bib:schlogl},
\[
2X + A \rightleftharpoons 3X,
\]
where $X$ is the reactant species and $A$ is a background species of
fixed density, can posess two stable equilibrium concentrations of the
species $X$.  In that case, the system can exhibit spontaneous pattern
formation as it breaks into domains of each concentration.  Because
kinetic fluctuations are important in the dynamics of such systems, it
is natural that lattice gas automata be applied to their study, and this
has been done with great success over the past five
years~\cite{bib:kap1,bib:kap2,bib:lawn,bib:dab}.

Reaction-diffusion lattice gas models typically allow reactant particles
to diffuse for some number of timesteps, $k$, between reactions.  The
diffusion steps obey SDB, while the reaction steps usually do not.  It
is remarkable that while natural chemically reacting systems seem to be
able to spontaneously generate patterns with microscopically reversible
laws of motion, all lattice gas models of such systems to date have
found it necessary to violate SDB.  There is no doubt that it is easier
to generate nontrivial structure in NSDB lattice gases.  Violations of
SDB can lead to the spontaneous generation of patterns and correlations,
and hence non-Gibbsian equilibria~\cite{bib:buss}.  In such situations,
however, the Boltzmann molecular chaos assumption is particularly
suspect, and the theoretical analysis of the system becomes difficult or
impossible.  Only in the limit of large $k$ has analytic progress been
made to date; at low $k$ the Boltzmann theory is known to be seriously
in error~\cite{bib:dab}.

In this paper, we describe a simple lattice gas model for Schl\"{o}gl's
second chemical reaction.  Because the reaction steps of this lattice
gas do not obey SDB, the equilibria are non-Gibbsian.  We derive a
self-consistent set of equations for the exact homogeneous equilibria
using cluster-expansion methods.  We solve these equations in the
two-particle BBGKY approximation; in this approximation these equations
are similar to those arising from the method recently developed by
Bussemaker et.\ al.~\cite{bib:bed}.  Comparing our results to numerical
experiment, we find that this approximation describes the equilibria far
more accurately than the Boltzmann approximation.  We also find,
however, that spurious solutions to the equilibrium equations appear
which can only be removed by including effects due to three-particle
correlations.  These spurious solutions are an important artifact of
this technique, and we argue that it is necessary to pay very close
attention to them in any such analysis.

\section{Description of the Schl\"{o}gl Model Lattice Gas}
\subsection{Schl\"{o}gl's Second Chemical Reaction}

Our starting point is the following generalization of Schl\"{o}gl's
second chemical reaction~\cite{bib:schlogl}:
\bgeas
2X + A & \stackrel{k_2^\pm}{\rightleftharpoons} & 3X \\
 X + B & \stackrel{k_1^\pm}{\rightleftharpoons} & 2X \\
     C & \stackrel{k_0^\pm}{\rightleftharpoons} &  X,
\eeas
where $X$ is the reactant species, $A$, $B$, and $C$ are background
species of fixed density, and the $k_j^\pm$ are the forward ($+$) and
reverse ($-$) rates for the reaction with $j$ reactant molecules on the
left.  Denoting the density of species $Y$ by $N_Y$, the stoichiometric
equation for this reaction is
\bgeas
\frac{dN_X}{dt}
   & = & k_2^+N_A N_X^2 - k_2^-N_X^3 +
         k_1^+N_B N_X   - k_1^-N_X^2 +
         k_0^+N_C       - k_0^-N_X \\
   & = & \kappa_0 - \kappa_1N_X + \kappa_2N_X^2 - \kappa_3N_X^3
\eeas
where we have defined the stochiometric coefficients,
\bgeas
\kappa_0 & = & k_0^+N_C \\
\kappa_1 & = & k_0^- - k_1^+N_B \\
\kappa_2 & = & k_2^+N_A - k_1^- \\
\kappa_3 & = & k_2^-.
\eeas
Finally, to model the stochastic motion of the reactant $X$ between
reactions, we add a diffusive term to obtain the reaction-diffusion
equation,
\bge
\frac{\partial N_X}{\partial t} = D\nabla^2 N_X +
   \kappa_0 - \kappa_1 N_X + \kappa_2 N_X^2 - \kappa_3 N_X^3.
\label{eq:sde}
\ee

\begin{figure}
\center{
\mbox{
\epsffile[45 140 180 270]{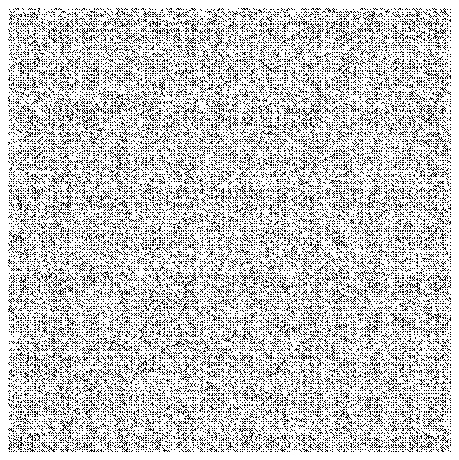}
\epsffile[45 140 180 270]{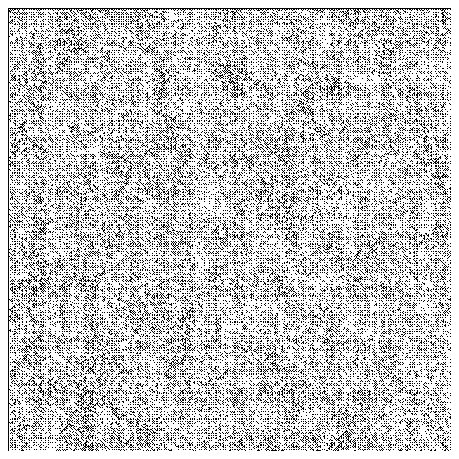}
\epsffile[45 140 180 270]{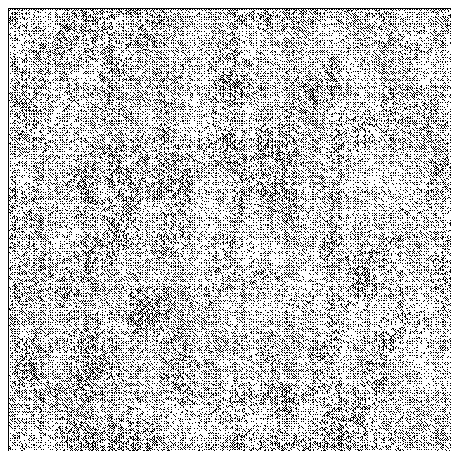}}}
\center{
\mbox{
\epsffile[45 140 180 270]{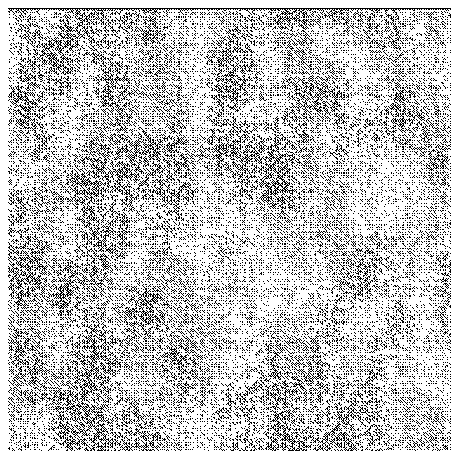}
\epsffile[45 140 180 270]{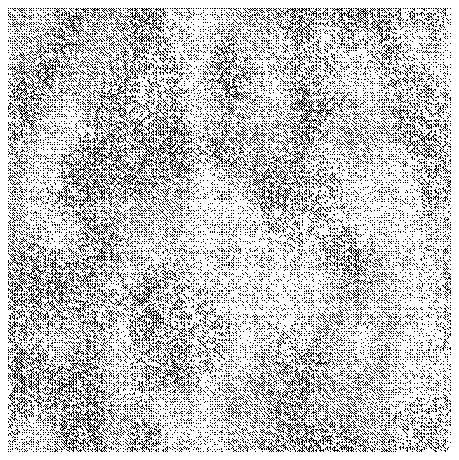}
\epsffile[45 140 180 270]{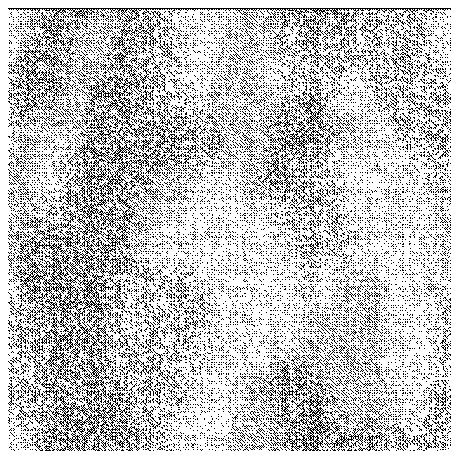}}}
\vspace{0.75truein}
\caption{{\bf Evolution of the Schl\"{o}gl model from random initial
conditions} yields domains of both low and high density, separated by
sharp gradients whose width is governed by the diffusive term in the
rate equation.}
\label{fig:evo}
\end{figure}

Note that Eq.~(\ref{eq:sde}) allows for up to three spatially uniform
equilibria, corresponding to the roots of the cubic.  When there are
three roots and $\kappa_3 > 0$, the low-density and high-density roots,
denoted by $N_X^-$ and $N_X^+$ respectively, are easily seen to be
stable to small fluctuations, while the middle root, $N_X^0$, is
unstable.  The evolution of Eq.~(\ref{eq:sde}) from generic initial
conditions thus yields domains of constant density $N_X^-$ and $N_X^+$,
separated by sharp gradients whose widths are governed by the diffusive
term in Eq.~(\ref{eq:sde}).  (See Fig.~\ref{fig:evo}.)

\subsection{Lattice Gas Model}

We model the kinetics of the generalized Schl\"{o}gl reaction by a
lattice gas automaton.  This consists of a regular lattice, ${\cal L}$,
with $n$ lattice vectors at each site; we denote the lattice vectors by
$\bfci$, where $i\in\{ 1,\ldots,n\}$.  The state of the system at time
$t$ is then completely specified by the quantities $\nixt\in\{ 0,1\}$
where $i\in\{ 1,\ldots,n\}$ and $\bfx\in {\cal L}$.  We have $\nixt=1$
if there is a particle with velocity $\bfci$ at position $\bfx$ at time
$t$, and $\nixt=0$ otherwise.

The evolution of the lattice gas for one timestep takes place in two
substeps.  In the {\it propagation} substep, the particles simply move
along their corresponding lattice vectors,
\[
\nsup{i}(\bfxci,\tpdt)\leftarrow\nixt.
\]
This is followed by the {\it collision} substep, in which the newly
arrived particles change their state.  The collisions are chosen to
model the reactive and diffusive dynamics of species $X$.  Their effect
is captured in the collision operator, $\omega^i$, which gives the
increase in the number of particles moving along direction $i$ due to
collisions.  In terms of this collision operator, the full equation of
evolution of the lattice gas may be written
\bge
\nsup{i}(\bfxci,\tpdt) = \nixt + \omega^i (\nsup{*}\bfxt),
\label{eq:evo}
\ee
where the dependence of $\omega^i$ on $n^*\bfxt$ indicates that each
component of the collision operator can depend on all the components
$\nsup{i}$ at the local site.

\begin{figure}
\center{
\epsffile{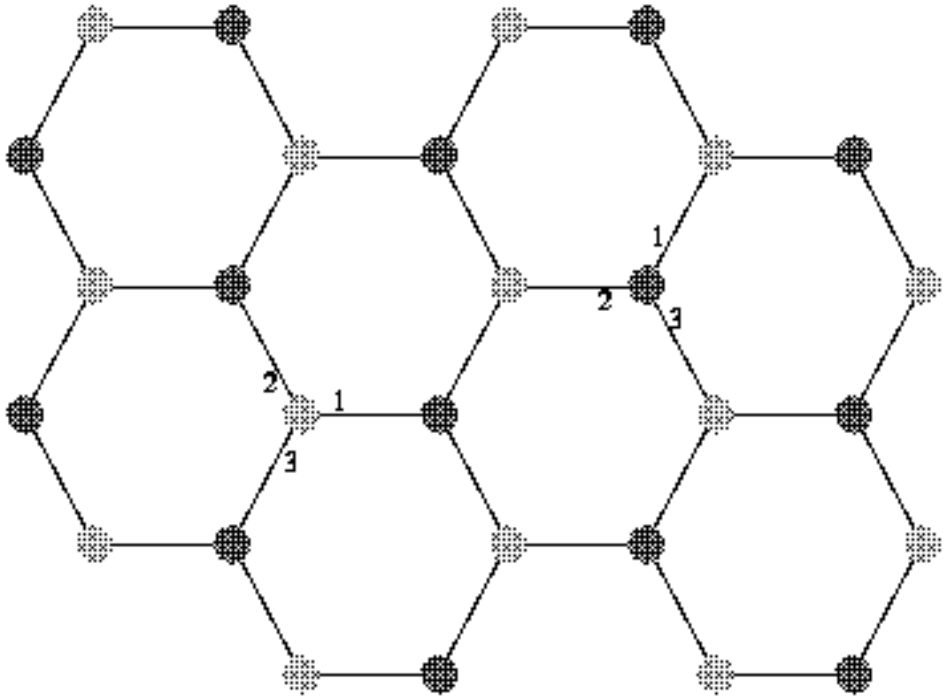}}
\vspace{0.75truein}
\caption{{\bf The hexagonal lattice}, with the checkerboard coloring and
the enumeration of the three bits at each site.}
\label{fig:lat}
\end{figure}

In this work, we restrict our attention to the Schl\"{o}gl model in two
dimensions.  We use a hexagonal (honeycomb) lattice because it has only
three bits of state at each site ($n=3$), thereby greatly simplifying
the analysis; at the same time, it is sufficiently symmetric to ensure
the isotropic form of the density balance equation, Eq.~(\ref{eq:sde}).
This lattice is illustrated in Fig.~(\ref{fig:lat}).  Note that such a
lattice can be colored like a checkerboard; note also that the
correspondence between the bits and the lattice vectors is rotated by
$\pi/3$ for the differently colored sites.

\subsection{The Collision Operator}

Following previous work on the modeling of chemical reactions by lattice
gases~\cite{bib:kap1,bib:kap2,bib:lawn,bib:dab}, we define two types of
interparticle collisions.  The chemical reactions take place in {\it
reactive collisions} in which particle number does not need to be
conserved.  Between reactions, the particles execute {\it diffusive
collisions} in which particle number is conserved.  Both types of
collision processes are {\it stochastic}; that is, the outgoing state of
a collision depends on one or more random bits that must be generated at
each site at each time step, as well as on the incoming state.  Reactive
collisions occur once every $k$ timesteps; the remainder of the
collisions are diffusive.

We need to carefully define the dynamics of the reactive and diffusive
collisions, and thence the form of the respective collision operators,
$\omega_R$ and $\omega_D$.  Because there are three bits per site, each
site can be in one of eight states.  We enumerate these states by
specifying the three bit values, i.e., $000, 001,\ldots 111$.  The
collision process can then be completely determined by specifying the
outgoing state corresponding to each incoming state.  Since the lattice
gas is stochastic, this specification may depend on one or more random
bits.

Let $a(s\rightarrow s')$ be $1$ if a collision takes state $s$ to state
$s'$, and $0$ otherwise.  Clearly, for each incoming state $s$,
$a(s\rightarrow s')$ can equal $1$ for exactly one $s'$, and must equal
$0$ for all others.  In terms of this {\it transition matrix}, the
collision operator can be written
\bge
\omega^i(\nsup{\ast}) =
   \sum_{s,s'} a(s\rightarrow s')
   (\supscrpt{s}{\prime i} - \supscrpt{s}{i})
   \prod_{j=1}^n \delta_{\nsup{j},\supscrpt{s}{j}},
\label{eq:collop}
\ee
where $\delta_{ij}\equiv 1-i-j+2ij$ is the Kronecker delta of the two
bits $i$ and $j$.  Together, Eqs.~(\ref{eq:evo}) and (\ref{eq:collop})
are a complete specification of the dynamics of the lattice gas.  Note
that $a(s\rightarrow s')$ may depend on random bits.

\subsection{The Boltzmann Equation}

We now suppose that we have prepared an ensemble of lattice gas
simulations, on grids of the same size, with initial conditions that are
sampled from some distribution.  We may take averages across this
ensemble.  Denoting $\nnixt\equiv\langle\nixt\rangle$, the ensemble
average of Eq.~(\ref{eq:evo}) is
\[
\nnsup{i}(\bfxci,\tpdt) = \nnixt + \langle\omega^i (\nsup{*}\bfxt)\rangle.
\]

We are hampered from taking the ensemble average of the collision
operator, Eq.~(\ref{eq:collop}), by the fact that it is generally a
nonlinear function of the $\nixt$, and the average of the product is not
equal to the product of the averages unless the quantities involved are
uncorrelated.  The simplest approximation to make is the {\it Boltzmann
molecular chaos} assumption that the particles entering a collision are
uncorrelated; in this case, the ensemble average of $\omega^i$ yields
the {\it Boltzmann collision operator},
\[
\Omega^i(\nnsup{\ast}) =
   \sum_{s,s'} A(s\rightarrow s')
   (\supscrpt{s}{\prime i} - \supscrpt{s}{i})
   \prod_{j=1}^n
      \left(\nnsup{j}\right)^{\supscrpt{s}{j}}
      \left(1-\nnsup{j}\right)^{1-\supscrpt{s}{j}},
\]
where $A(s\rightarrow s')\equiv\langle a(s\rightarrow s')\rangle\in
[0,1]$ is the {\it ensemble-averaged transition matrix}.

Note that there are three one-particle states ($001$, $010$, $100$),
three two-particle states ($110$, $101$, $011$), one zero-particle state
($000$), and one three-particle state ($111$).  Let $|s|$ denote the
number of particles in state $s$, so for example $|101|=2$.  For the
lattice gas considered here, the mean outcome of both diffusive and
reactive collisions depends only on the total number of incoming
particles, and is always uniformly distributed over the states of the
outgoing particle number.  Mathematically, this means that the
$A(s\rightarrow s')$ can depend only on $|s|$ and $|s'|$, and can thus
be tabulated as in Table~\ref{tab:eatm}, where $\mixten{P}{i}{j}$ is the
probability that a collision will take a state with $j$ particles into a
state with $i$ particles.  The evolution equation in this approximation
is the {\it Boltzmann equation},
\bge
\nnsup{i}(\bfxci,\tpdt) = \nnixt + \Omega^i (\nsup{*}\bfxt).
\label{eq:eaevo}
\ee

For the diffusive collisions, we must have
\[
\mixten{P}{i}{j} = \mixten{\delta}{i}{j},
\]
where $\mixten{\delta}{i}{j}$ is the Kronecker delta.  Thus, a diffusive
collision is nothing more than a random permutation of the three
incoming bits.  Calculation of the corresponding Boltzmann collision
operator is straightforward yielding
\bge
\Omega_D^i (N^*) = -\frac{2}{3}N^i +\frac{1}{3}N^{i+1} +\frac{1}{3}N^{i+2},
\label{eq:eacollopd}
\ee
where the superscript of $N$ is understood to be taken modulo 3.

To simplify the algebra for the reaction step, we henceforth restrict
our attention to the following specific values for the particle
transition probabilities,
\[
\mixten{P}{i}{j} =
   \mixten{
   \left(
   \begin{array}{cccc}
   2/3 & 1/3 &   0 &   0 \\
   2/3 & 1/3 &   0 &   0 \\
     0 &   0 & 1/3 & 2/3 \\
     0 &   0 & 1/3 & 2/3
   \end{array}
   \right)}{i}{j},
\]
for $i,j\in\{ 0,1,2,3\}$.  Calculation of the corresponding Boltzmann
collision operator yields
\bge
\Omega_R^i (N^*) =
   \frac{1}{9} - N^i +
   \frac{7}{9} \left(N^0 N^1 + N^0 N^2 + N^1 N^2\right) -
   \frac{14}{9} N^0 N^1 N^2.
\label{eq:eacollopr}
\ee
A complete Boltzmann description of the system is given by
Eq.~(\ref{eq:eaevo}), using Eq.~(\ref{eq:eacollopr}) once every $k$
timesteps and Eq.~(\ref{eq:eacollopd}) otherwise.

\subsection{Boltzmann Equilibria}

Note that the Boltzmann equation, Eq.~(\ref{eq:eaevo}), admits
homogeneous, isotropic equilibria, $N^0=N^1=N^2=f$, where $f$ obeys
$\Omega (f)=0$.  Note also that the diffusive collision operator,
Eq.~(\ref{eq:eacollopd}), satisfies $\Omega_D (f)=0$ identically.  We
thus find homogeneous, isotropic equilibria by demanding that the
reaction step do likewise,
\bgea
0 & = & \Omega_R^i (f) \nonumber \\
  & = & \frac{1}{9} - f + \frac{7}{3} f^2 - \frac{14}{9} f^3 \nonumber \\
  & = & \frac{1}{9} (1-2f) (7f^2-7f+1). \label{eq:beq}
\eea
This has roots at $f = \frac{1}{2}$ and $f = \frac{1}{2}\left(1\pm
\sqrt{\frac{3}{7}}\right)$.  Fig.~(\ref{fig:evo}) displays the evolution of
the lattice gas model for these parameters, with the initial condition
$f=\frac{1}{2}$ everywhere.

\section{Exact Equations of Motion}

The exact microscopic equations of motion for any lattice gas are easily
described in terms of the multi-particle means $N^\alpha$ (following the
notation of our previous paper~\cite{bib:long} we denote by $\alpha$ an
arbitrary subset of the bits (particles) in the system, and by
$N^\alpha$ the ensemble average of the product of those bits).  In terms
of these means, the exact time-development equation is
\bge
\nnsup{\alpha}(\tpdt) =
   \advop{\alpha}{\beta} \kop{\beta}{\gamma} \nnsup{\gamma} (t),
   \label{eq:exactm}
\ee
where we use the  convention of summing over any index which appears
twice on one side of an equation and not at all on the other side.
In this equation, $\advop{\alpha}{\beta}$ is an advection operator,
described by a permutation matrix on the set of bit sets $\alpha$,
which carries each bit of the system forward along its associated
velocity vector.  The operator
$\kop{\beta}{\gamma}$ describes the collision process.  It can be
factorized into contributions from each lattice site,
\bge
   \kop{\beta}{\gamma}
   = \prod_{\bfx \in L_\beta} \vop{\beta_\bfx}{\gamma_\bfx},
\ee
where $L_\beta$ is the set of vertices associated with bits in $\beta$
and $\beta_\bfx$ is the set of bits in $\beta$ at the lattice site
$\bfx$.  The mean vertex coefficients $\vop{\beta}{\gamma}$ are
related to the state transition probabilities $A (s \rightarrow s')$ through
\begin{equation}
\mixten{V}{\mu}{\nu} = \sum_{s' \supseteq \mu}
\sum_{s \subseteq \nu}(-1)^{| \nu | - | s |} A (s \rightarrow s').
\end{equation}

The exact time-development equation (\ref{eq:exactm}) can be rewritten
in terms of connected correlation functions (CCF's) using the standard
cluster expansion.  The means are expressed in terms of the CCF's
through
\bge
   \nnsup{\alpha} =
   \fsup{\alpha}(\ggamsup{\ast}) =
   \sum_{\zeta \in \pi(\alpha)}
   \ggamsup{\zeta_1} \ggamsup{\zeta_2} \ldots \ggamsup{\zeta_q},
   \label{eq:meanccf}
\ee
where $\pi(\alpha)$ is the set of all partitions of $\alpha$ into
disjoint subsets, $\zeta_1, \ldots, \zeta_q$.  For example, we have
$\nnsup{a} = \ggamsup{a}$, $\nnsup{ab} = \ggamsup{ab} + \ggamsup{a}
\ggamsup{b}$.  This relation can be inverted to express the CCF's in
terms of the means, $\ggamsup{\alpha} = \gsup{\alpha}(\nnsup{\ast})$.

We can now rewrite (\ref{eq:exactm}) as
\begin{equation}
   \ggamsup{\alpha}(\tpdt) =
   \advop{\alpha}{\beta} \gsup{\beta}(\kop{\ast}{\gamma}
   \fsup{\gamma}(\ggamsup{\ast})).
\label{eq:exactc}
\end{equation}
This exact equation has been used as a starting point in previous
works~\cite{bib:long,bib:bed}.  It has been applied to SDB lattice
gases~\cite{bib:long}, where the equilibria have no correlations and the
expression on the right hand side can be linearized in terms of the
CCF's $\Gamma^\alpha$ with $| \alpha | \geq 2$.  Eq.~(\ref{eq:exactc})
has also been applied to NSDB lattice gases by Bussemaker et
al.~\cite{bib:bed}\ who neglected CCF's $\Gamma^\alpha$ with $| \alpha |
\geq 3$, and thereby derived the 2-particle BBGKY equations for NSDB
lattice gases.

It has been shown~\cite{bib:long} that the linearized form of
(\ref{eq:exactc}) can naturally be expressed in terms of a sum over
diagrams, each of which is weighted by a product of factors associated
with each vertex at each time step.  There are a finite number of
possible vertices, so that a complete formulation of the dynamics of a
SDB lattice gas can be given in terms of a set of ``Feynman rules'' for
allowed diagrams and vertex weights.

An analogous diagrammatic description can be given for the exact
nonlinear equations (\ref{eq:exactc}).  The nonlinear diagrammatic
expansion can be derived by proving a general factorization theorem for
the time development of CCF's including particles at different vertices.
The essential ingredient in proving this factorization is the
observation that if a set of variables $ \alpha$ depends stochastically
on another set of variables $ \beta$, so that the CCF $\Gamma^\alpha$ is
given by
\begin{equation}
\Gamma^\alpha =\mixten{{\cal K}}{\alpha}{ \xi}
\prod_{ \xi_i \in  \xi} \Gamma^{\xi_i}
\end{equation}
where $\xi =\{\xi_1, \ldots, \xi_m\}$ is a set of (not necessarily
disjoint) subsets of $\beta$, then the CCF of $\alpha$ joined with a set
of variables $\gamma$ which are not dependent on $\beta$ is given by
\begin{equation}
\Gamma^{ \alpha \cup \gamma} = \mixten{{\cal K}}{\alpha}{ \xi}
\sum_{\zeta \in \pi_m(\gamma)}
\prod_{i} \Gamma^{ \xi_i \cup \zeta_i}
\label{eq:split}
\end{equation}
where $\zeta=\{\zeta_1, \ldots, \zeta_m\}$ is summed over all
partitions of $\gamma$ into precisely $m$ distinct sets.
This result essentially states that once we know  an
expression for the outgoing CCF's at a particular vertex of a lattice
gas in terms of the incoming CCF's, we can calculate the CCF of a set
of particles at multiple lattice sites by applying (\ref{eq:split}) at
each vertex separately.
The general expression for an outgoing CCF at one vertex can be
written by expanding
\[
   \phisup{\beta}(\ggamsup{\ast}) \equiv
   \gsup{\beta}(\kop{\ast}{\gamma} \fsup{\gamma}(\ggamsup{\ast})).
\]
as an explicit polynomial in the CCF's; i.e.,
\begin{equation}
   \phisup{\beta}(\ggamsup{\ast})  =
\mixten{{\cal K}}{\beta}{  \xi} \prod_{\xi_i \in \xi}
\Gamma^{\xi_i}
\end{equation}
where $\xi =\{ \xi_1, \ldots, \xi_k\}$ is summed over all sets of
CCF's with nonzero coefficients.
Each time the equation (\ref{eq:split}) is applied at a
particular vertex, the correlated quantities at the other vertices are
carried along and divided up in all possible ways among the incoming
CCF's.   A simple example of this result is that when $a$ is an
outgoing particle from  a vertex with incoming particles
$b_1,b_2,b_3$, and $c$ is an outgoing particle from a different vertex
at the same time step, we have (for a general lattice gas)
\[
\Gamma^a = f (\{\Gamma^{b_i}\}, \{\Gamma^{b_ib_j}:i \neq j\},
\Gamma^{b_1b_2b_3})
\]
and
\begin{equation}
\Gamma^{ac} = \sum_{i} \frac{\partial f}{ \partial \Gamma^{b_i}}
\Gamma^{b_i c}+ \frac{1}{2}\sum_{i \neq j} \frac{\partial f}{ \partial
\Gamma^{b_i b_j}}\Gamma^{b_i b_j c} + \frac{\partial f}{ \partial
\Gamma^{b_1b_2b_3}}  \Gamma^{b_1b_2b_3 c}.
\label{eq:simplesplit}
\end{equation}
The proof of (\ref{eq:split}) follows fairly easily by induction on $j$
and $k$.  The details of this proof and the general factorization
theorem in the nonlinear case will be given in a separate
publication~\cite{bib:btnext}.  The result (\ref{eq:simplesplit}), which
follows directly from (\ref{eq:meanccf}) will be sufficient for our
purposes in this paper.

We conclude this section with a derivation of a simple form of the
factorization theorem which we will need in the sequel.  Assume that
at one vertex we have an outgoing particle $A$ and incoming particles
$a,b,c$, and that at another vertex we have an outgoing particle
$\bar{A}$ and incoming particles $\bar{a}, \bar{b}, \bar{c}$.  We wish
to find the dependence of the outgoing CCF $\Gamma^{A \bar{A}}$ on the
incoming correlations, neglecting all CCF's between 3 or more
particles.  It will suffice for us to know the dependence of the
outgoing 1-particle means on the incoming 1- and 2- particle CCF's at
each vertex.  Thus, we can write
\[
\Gamma^{A} = f (\Gamma^a, \Gamma^b, \Gamma^c,
\Gamma^{ab}, \Gamma^{bc}, \Gamma^{ac}) + {\cal O}(C_3)
\]
and
\[
\Gamma^{\bar{A}} = g (\Gamma^{\bar{a}}, \Gamma^{\bar{b}},
\Gamma^{\bar{c}},
\Gamma^{\bar{a} \bar{b}}, \Gamma^{\bar{b} \bar{c}}, \Gamma^{\bar{a}
\bar{c}}) + {\cal O}(C_3)
\]
where by ${\cal O}(C_i)$ we denote quantities dependent on CCF's of $i$
or more variables.  Applying (\ref{eq:simplesplit}) once, we have
\[
\Gamma^{A \bar{A}} =
\frac{\partial f}{\partial \Gamma^a}
\Gamma^{a \bar{A}} +
\frac{\partial f}{\partial \Gamma^b}
\Gamma^{b \bar{A}} +
\frac{\partial f}{\partial \Gamma^c}
\Gamma^{c \bar{A}} +{\cal O}(C_3).
\]
Applying (\ref{eq:simplesplit}) again, we have
\begin{equation}
\Gamma^{A \bar{A}} =
\frac{\partial f}{\partial \Gamma^\alpha}
\frac{\partial g}{\partial \Gamma^{\bar{\alpha}}}   \Gamma^{\alpha
\bar{\alpha}}
+ {\cal O}(C_3),
\label{eq:simplefactor}
\end{equation}
where $\alpha, \bar{\alpha}$ are summed over $\{ a,b,c\}$ and
$\{\bar{a}, \bar{b}, \bar{c}\}$ respectively.  Note that this equation
has a diagrammatic interpretation because the coefficient associated
with the propagation of a pair of correlated quantities at different
vertices factorizes into contributions from each vertex separately.  We
will use this simple factorization result in the next section to compute
the exact 2-particle BBGKY equations for the equilibria of the Schl\"ogl
model lattice gas.

\section{Exact Equilibria of Schl\"{o}gl model}

We will now consider the exact equations of motion for the Schl\"{o}gl
model lattice gas defined in Section 2.  By neglecting correlations
between more than two particles, we arrive at the 2-particle BBGKY
equations, which we then solve using the diagrammatic method.  The
2-particle BBGKY equations were described for a general NSDB lattice gas
by Bussemaker et al.~\cite{bib:bed}, who gave an iterative method for
finding solutions to these equations.  Although the equations we are
solving here are essentially equivalent to those which would be found by
applying the methods of these authors to the Schl\"{o}gl model lattice
gas, our diagrammatic method of solution of these equations is rather
different.  Using the diagrammatic formalism, there is no issue of
convergence as there is with the iterative method; furthermore, in our
analysis, there is no question of uniqueness of solutions -- we can
identify directly all distinct solutions of the 2-particle equations.
In fact, we find that the 2-particle BBGKY equations have spurious
solutions for the lattice gas considered here.

The first step in writing the exact equations for the Schl\"{o}gl model
lattice gas is to write the exact equation for CCF's at a single vertex.
There are two sets of such equations, corresponding to the diffusive and
reactive vertices, respectively.  The mean vertex coefficients
$\vop{\alpha}{\beta}$ for both of these vertex types are symmetric with
respect to permutations of incoming and outgoing bits separately, and
therefore are only functions of the numbers of bits in $\alpha$ and
$\beta$.  These vertex coefficients are easily calculated and are
tabulated in Tables~\ref{tab:vcdv} and \ref{tab:vcrv}.

{}From these vertex coefficients, we can use (\ref{eq:exactc}) to write
the exact equations for the outgoing CCF's from a diffusive or reactive
vertex in terms of the incoming CCF's.  These equations are again
invariant under arbitrary independent permutations of the incoming and
outgoing bits.  Labeling the outgoing particles by $A,B,C$ and the
incoming particles by $a,b,c$, the equations for a diffusive vertex are
given by
\begin{eqnarray}
\Gamma^A & =  &\frac{1}{3}
(\Gamma^a + \Gamma^b + \Gamma^c)\nonumber\\
\Gamma^{A B} &=  &\frac{1}{3}
(\Gamma^{a b} + \Gamma^{a c} + \Gamma^{b c})  \label{eq:exactdiffusive}\\
\Gamma^{A B C}   &=  & \Gamma^{a b c}\nonumber.
\end{eqnarray}
The 1-particle equation for a reactive vertex is
\bgea
\Gamma^{A} &=&
\frac{1}{9} +\frac{7}{9}(  \Gamma^{a} \Gamma^{b} +
 \Gamma^{a} \Gamma^{c} +  \Gamma^{b} \Gamma^{c} -
   2 \Gamma^{a} \Gamma^{b} \Gamma^{c} +  \Gamma^{a b} -
2 \Gamma^{c} \Gamma^{a b} \\
 & & + \Gamma^{a c} - 2 \Gamma^{b} \Gamma^{a c} +
 \Gamma^{b c} -
  2 \Gamma^{a} \Gamma^{b c} - 2 \Gamma^{a b c}).
\label{eq:1particle}
\eea
The equations for 2- and 3-particle outgoing CCF's are straightforward
to calculate but are algebraically more complicated than
Eq.~(\ref{eq:1particle}).  Note that setting the two- and three-particle
correlations to zero in this equation, and setting all 1-particle
correlations to the mean occupation number $f = \Gamma^a = \Gamma^b =
\Gamma^c$, reproduces the Boltzmann equilibrium, Eq.~(\ref{eq:beq}).

Henceforth, we will restrict attention to uniform equilibria, so that
the correlations are independent of spatial coordinate or orientation.
We denote the equilibrium values of the 1-, 2-, and 3-particle CCF's
{\em entering} a reactive vertex by $I_1$, $I_2$, and $I_3$
respectively.  Similarly, we denote the CCF's {\em leaving} a reactive
vertex by $O_1$, $O_2$, and $O_3$.  The exact equations of motion for
the 1- and 2-particle CCF's leaving a reactive vertex are
\begin{eqnarray}
O_1 & = &
\frac{1}{9} + \frac{7 I_1^2}{3} - \frac{14 I_1^3}{9} + \frac{7 I_2}{3} -
\frac{14 I_1 I_2}{3} -\frac{14 I_3}{9} \nonumber
 \\
O_2 & = &
\frac{-1}{81} + \frac{49 I_1^2}{27} - \frac{98 I_1^3}{81} -
\frac{49 I_1^4}{9} + \frac{196 I_1^5}{27} -
  \frac{196 I_1^6}{81} + \frac{49 I_2}{27} - \frac{98 I_1 I_2}{27}
 -\frac{98 I_1^2 I_2}{9} \label{eq:equilibrium}\\
& &
+ \frac{784 I_1^3 I_2}{27} - \frac{392 I_1^4 I_2}{27} - \frac{49 I_2^2}{9}
+ \frac{196 I_1 I_2^2}{9} -
  \frac{196 I_1^2 I_2^2}{9}
 - \frac{98 I_3}{81} + \frac{196 I_1^2 I_3}{27} - \frac{392 I_1^3 I_3}{81}
\nonumber \\ & & +
  \frac{196 I_2 I_3}{27} - \frac{392 I_1 I_2 I_3}{27} - \frac{196 I_3^2}{81}
\nonumber
\end{eqnarray}
The equation for $O_3$ can be similarly written, but is slightly more
complicated and will not be used here.  Recall that, as was demonstrated
in the previous section, the exact dynamical equation of an arbitrary
number of correlated quantities can be described in terms of the exact
equations for the CCF's at a single vertex.  Thus,
Eq.~(\ref{eq:equilibrium}), along with the corresponding equation for
$O_3$, gives a complete description of the equations of motion of {\it
all} CCF's at a reactive timestep.

To complete the equilibrium equations (\ref{eq:equilibrium}), we must
determine the relations between the outgoing correlations $O_i$ from a
reactive vertex and the incoming correlations $I_i$.  Referring back to
Eqs.~(\ref{eq:exactc}) and (\ref{eq:exactdiffusive}), we see that at
diffusive time steps, the correlations essentially perform random walks
on the honeycomb lattice.  Thus, the correlation $I_1$ entering a fixed
reactive vertex at some time step is a weighted sum of outgoing
correlations $O_1$ from vertices at the previous reactive timestep, with
total weight 1.  Since we have assumed an isotropic equilibrium, we have
an equilibrium density $f$ satisfying
\begin{equation}
f =I_1 = O_1.
\end{equation}
It is interesting to note that by using this equality in the
first equation in (\ref{eq:equilibrium}) we can write an exact
expression for $I_3$ in terms of $I_2$ and $I_1$.  Inserting this
expression into the second equation of (\ref{eq:equilibrium}), we find
that the terms in $I_2$ cancel and we have the result
\begin{equation}
O_2 = -\frac{1}{9}  + f -f^2
\label{eq:exactidentity}
\end{equation}
Note that this equation is {\em exact}, and must be satisfied by any
isotropic equilibrium of the system.

In principle, we would now like to find an exact set of expressions
relating the quantities $I_2,I_3$ to outgoing quantities $O_2,O_3$ by
iterating the exact equations of motion.  However, this is technically
infeasible since such a calculation would involve a sum over diagrams
involving arbitrary numbers of correlated quantities.  Thus, we shall
now restrict to the 2-particle BBGKY equations by neglecting
correlations of more than 2 particles.  By making this simplification,
we derive a simple set of equations whose solutions give the equilibria
of the lattice gas in the 2-particle BBGKY approximation.

Neglecting 3-particle correlations, and setting $f = O_1 =I_1$,
the exact equations for the CCF's at a reactive vertex become
\begin{eqnarray}
f & = &
\frac{1}{9} + \frac{7 f^2}{3} - \frac{14 f^3}{9} + \frac{7 I_2}{3} -
\frac{14 f I_2}{3} \nonumber\\
& = & f + \frac{1}{9} (1-2 f) (1-7 f + 7 f^2 + 21 I_2)
 \label{eq:2equilibrium1}
\end{eqnarray}
\begin{eqnarray}
O_2 & = &
\frac{-1}{81} + \frac{49 f^2}{27} - \frac{98 f^3}{81} -
\frac{49 f^4}{9} + \frac{196 f^5}{27} -
  \frac{196 f^6}{81} + \frac{49 I_2}{27} \nonumber\\
& & - \frac{98 f I_2}{27} -
\frac{98 f^2 I_2}{9} +
  \frac{784 f^3 I_2}{27} - \frac{392 f^4 I_2}{27} - \frac{49 I_2^2}{9}
+ \frac{196 f I_2^2}{9} -
  \frac{196 f^2 I_2^2}{9}. \label{eq:2equilibrium2}
\end{eqnarray}
The first of these equations is satisfied whenever either
\[
f = \frac{1}{2}
\]
or
\begin{equation}
I_2
= -\frac{1}{21}  (1-7 f + 7 f^2)
\label{eq:i2}
\end{equation}
The solution $f = 1/2$ corresponds to the unstable equilibrium of
the Boltzmann theory, and shows that this unstable equilibrium still
exists in the 2-particle BBGKY approximation.  We will not discuss
this solution further here.  Inserting (\ref{eq:i2}) into
(\ref{eq:2equilibrium2}), we again derive the identity
(\ref{eq:exactidentity}), so this identity still holds in the
2-particle BBGKY approximation.

To find all solutions to the 2-particle BBGKY equilibrium equations, it
remains for us to find a relation between $I_2$ and $O_2$.  The analysis
of the flow of 2-particle correlations is slightly more subtle than that
of the 1-particle density.  Tracing back a given incoming correlation
$I_2$ to the previous reactive vertex ($k$ timesteps earlier), we find
that with some probability $\phi_k (1)$ the random walks of the
correlated quantities lead back to a pair of outgoing particles from a
single vertex associated with an outgoing correlation $O_2$.  However,
the remaining random walks (with probability $1-\phi_k (1)$) lead to a
pair of correlated quantities at different vertices.  For a fixed pair
of vertices, we denote such an outgoing correlation from a reactive
vertex by $O_{1,1}$.  Using Eqs.~(\ref{eq:1particle}) and
(\ref{eq:simplefactor}), we can expand $O_{1,1}$ in terms of incoming
CCF's of the 6 particles associated with the two vertices in question.
Making the 2-particle BBGKY approximation, we have
\begin{equation}
O_{1,1} = \lambda^2 (\sum_{{\rm pairs}} I_{1,1})
\end{equation}
where the sum is taken over all 9 possible pairs of incoming
particles, 1 from each vertex, which may be correlated, and where
\begin{equation}
\lambda =\lambda (f,I_2) = \frac{14}{9}  (f - f^2 -I_2)
=\frac{2}{ 27}  (1 + 14 f -14 f^2).
\label{eq:lambda}
\end{equation}
Note that $\lambda \leq 1/3$, with equality only when $f = 1/2$.  We can
repeat the above steps for the particles correlated in each term
$I_{1,1}$.  Moving back through $k-1$ diffusive vertices, associated
with random walks of the correlated quantities, we again have some set
of diagrams where the correlation originates in a pair of outgoing
particles from a single previous reactive vertex, and some other set of
diagrams where the correlated quantities are still separate.  Repeating
this analysis indefinitely, we find that the equilibrium correlations
$I_2$ and $O_2$ can be related by
\begin{equation}
I_2 = [\sum_{t = 1}^{ \infty}
\phi_k (t) (3 \lambda)^{2t-2}] O_2,
\label{eq:equilibrium3}
\end{equation}
where $\phi_k (t)$ is the weighted sum over all diagrams describing
random walks of 2 particles for $kt$ time steps on the honeycomb
lattice, where the particles leave a particular vertex on the first step
in a fixed pair of directions and arrive together at some possibly
different vertex at the final step.  In these diagrams, the particles
are not allowed to visit the same vertex at any time step divisible by
$k$ (reactive vertices), and when they visit the same vertex at any
other time step (diffusive vertices), they exit in different directions
with each possible pair of outgoing directions having equal probability
(corresponding to (\ref{eq:exactdiffusive})).  Note that the factor of 3
appears because the usual probability 1/3 of a given random bounce is
replaced by the weight $\lambda$.

\begin{figure}
\epsffile{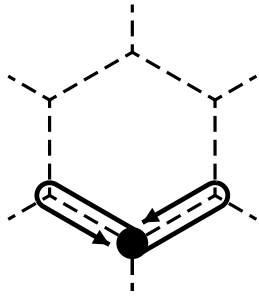}
\caption{{\bf Unique diagram} contributing to $\phi_2 (1)$.}
\label{fig:figure}
\end{figure}

As an example of a coefficient $\phi_k (t)$, it is easy to calculate
\[
\phi_2 (1) = \frac{1}{9},
\]
since the unique diagram which contributes is as shown in
Figure~\ref{fig:figure}.  Similarly, since at $t=2$ there are 30
diagrams which each contribute $(1/3)^6$, one finds that
\[
\phi_2 (2) = 30\left(\frac{1}{3}\right)^6 = \frac{10}{243}.
\]
It follows immediately from the random walk interpretation of $\phi_k
(t)$ that
\[
\sum_{t = 1}^{ \infty}  \phi_k (t) = 1,
\]
since the probability that two random walkers in 2D will eventually
collide is 1.  An immediate consequence is that the series
\[
\sum_{t = 1}^{ \infty}  \phi_k (t) (3 \lambda)^{2t-2}
\]
converges whenever $\lambda \leq 1/3$.  Furthermore, for $\lambda$
satisfying this condition, we can calculate the above series to
arbitrary accuracy; given any $\epsilon$, when $\lambda \leq 1/3$ we
can choose $T$ such that $\sum_{t = 1}^{T} \phi_k (t) > 1-\epsilon $,
and it follows immediately that
\[
\sum_{t  > T}   \phi_k (t) (3 \lambda)^{2t-2}< \epsilon.
\]
Thus, to calculate the sum to within an accuracy of $\epsilon$ we need
only calculate a finite number of coefficients $\phi_k (t)$, a task
which is easily performed numerically by a computer.

\begin{figure}
\epsfxsize=400pt
\epsffile{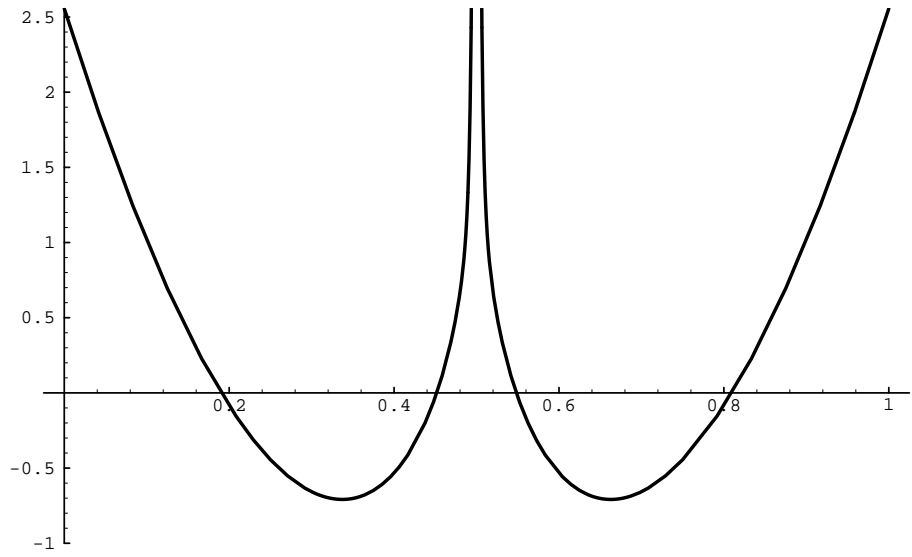}
\vspace{0.75truein}
\caption{{\bf Plot of $\zeta (f)$ versus $f$} for $k=3$.}
\label{fig:zeta}
\end{figure}

We may now use (\ref{eq:exactidentity}), (\ref{eq:i2}),  and
(\ref{eq:equilibrium3}) to derive a single equation
for the 2-particle BBGKY equilibrium density $f$,
\begin{equation}
\zeta (f)=3 (1-7 f + 7 f^2)  -7 (1-9 f + 9 f^2) \alpha
(f)= 0,
\label{eq:final}
\end{equation}
where
\begin{equation}
\alpha (f) = \sum_{t = 1}^{ \infty}  \phi_k (t)\left[\frac{2}{9}
(1 + 14 f -14 f^2) \right]^{2t-2}.
\label{eq:alpha}
\end{equation}
Since, as mentioned above, we can calculate $\alpha (f)$ to an arbitrary
degree of accuracy, it is a straightforward process to numerically
determine the values of $f$ which satisfy Eq.~(\ref{eq:final}) to an
arbitrary degree of accuracy.  We have performed such a numerical
analysis for $k$ ranging from 2 to 7.  For each value of $k$, we find
not two, but {\em four} distinct equilibria satisfying $\zeta (f)= 0$.
As an example, we graph in Figure~\ref{fig:zeta} the function $\zeta
(f)$ for $k = 3$.  This function has two zeros at $f\approx 0.1903$ and
$f\approx 0.8097$, which presumably correspond to the actual equilibria
of the system.  We will refer to these zeros as the ``primary''
solutions.  In addition, however, the function has two zeros near $f =
1/2$, which we will call the ``secondary'' solutions.  Because the
series for $\alpha (f)$ converges very slowly in the vicinity of $f =
1/2$, one might be suspicious of the secondary solutions.  To see that
such solutions must exist, however, we can observe that at $f = 1/2$ we
have $\alpha (1/2)= 1$ and therefore $\zeta (1/2)= 13/2$ for any $k$.
Since the function converges nicely and is negative above (below) the
lower (upper) primary solution, there must be a secondary pair of
solutions, just as we see in the graph.

\begin{figure}
\epsfxsize=400pt
\epsffile{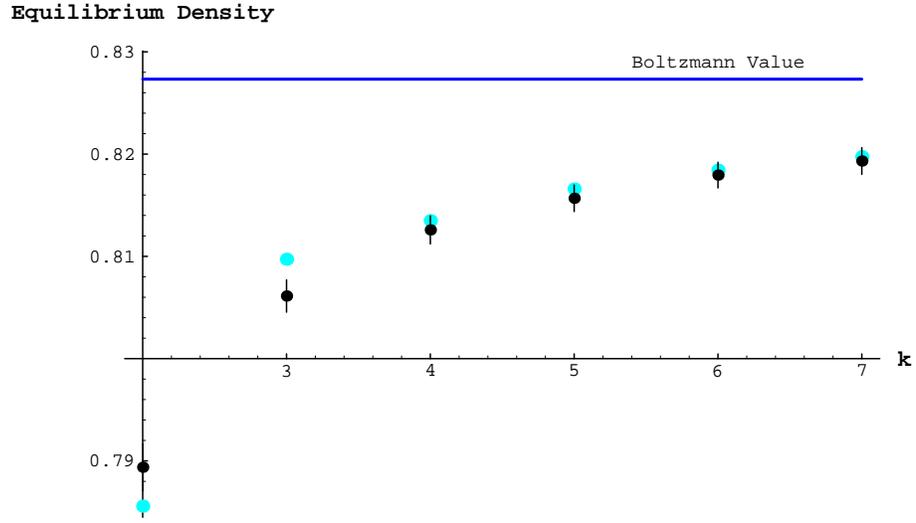}
\vspace{0.75truein}
\caption{{\bf Equilibrium density} versus $k$.  The black points with the
error bars are from numerical experiment, the gray points without error
bars are from the 2-particle BBGKY theory, and the line across the top
is the Boltzmann value.}
\label{fig:neq}
\end{figure}

Comparison of the primary equilibrium solutions with numerical results
{}from simulations of the lattice gas with various values of $k$ shows
that these solutions of the 2-particle BBGKY equations predict the exact
equilibria of the lattice gas system remarkably well.  This comparison
is given in Fig.~\ref{fig:neq}.
We see that the 2-particle BBGKY approximation gives an excellent
numerical prediction of the equilibria of the Schl\"{o}gl model lattice
gas.  However, the existence of the spurious secondary equilibria
demonstrates emphatically that one must be very careful when dealing
with truncations of the exact equations for a lattice gas.  In the work
of Bussemaker et al., for instance, an iterative method is used to solve
the 2-particle BBGKY equations~\cite{bib:bed}.  This approach can result
in a spurious equilibrium, with no indication that any other solution
exists.  Thus, without some further criterion for judging the validity
of a solution to these equations, it is difficult to evaluate the
results of such an analysis.

We will now proceed to give some simple analytic arguments which show
that the secondary solutions are highly sensitive to the introduction of
3-particle CCF's, and thus that they are suspect from a priori grounds.
First, let us observe that the introduction of a small amount of
3-particle correlation in $I_3$ would change (\ref{eq:i2}), which would
then read
\begin{equation}
I_2
= -\frac{1}{21}  (1-7 f + 7 f^2 ) + \frac{2I_3}{3 (1-2 f)}.
\end{equation}
If the correlation $I_3$ were small, this would cause a change in $I_2$
which would be small except in the region $f\approx 1/2$, where the
change would be dramatic.  A change in $I_2$ would in turn cause a
comparable change in $\lambda$ through (\ref{eq:lambda}).  Since the sum
(\ref{eq:alpha}) converges slowly in the region of $\lambda\approx 1/3$,
the value of $\alpha (f)$ is highly sensitive to a slight change in
$\lambda$ in this region, which is precisely the region where $f\approx
1/2$.  In fact, only a small change in $\lambda$ is needed to lower
$\alpha$ sufficiently that $\zeta (1/2)< 0$, which would result in a
disappearance of the spurious equilibria.

The composition of the two extreme sensitivities described here makes it
clear that the existence of the spurious equilibria are highly dependent
upon the vanishing of the 3-particle CCF $I_3$.  In fact, we have
extended our analysis to include a simple class of 3-particle diagrams
and found that with this minor modification, the spurious equilibria
completely disappear.  Specifically, one can take the exact 3-particle
equations at a vertex, and solve using the additional condition that
$I_3= \mu O_3$ where $\mu$ is the weight of some simple class of
diagrams involving 3 correlated particles.  For the case $k = 2$, the
simplest 3-particle diagram is the one where 3 particles leave a vertex,
and bounce directly back on the subsequent advective step.  This diagram
gives $\mu = 1/27$.  Exactly solving the resulting equations for the 1-,
2- and 3-particle CCF's, we find that there are precisely 2 solutions
(aside from the unstable solution at $f = 1/2$).  Thus, it seems clear
that the secondary equilibria generated by the 2-particle BBGKY
equations are spurious, since they can be removed by such a simple
perturbation.  Unfortunately, including an arbitrary set of 3-particle
diagrams, without performing the systematic 3-particle BBGKY
approximation, tends to reduce the effectiveness of the approximation;
thus, although the spurious equilibria are removed, the analysis
described here does not give more accurate predictions for the actual
equilibria than the 2-particle BBGKY analysis.  To have a significantly
improved approximation to the actual equilibria of the lattice gas, one
would need to use a more complicated approximation scheme such as the
complete 3-particle BBGKY approximation.

We conclude this section with a brief discussion of finite size effects.
For any finite lattice, the complete equations of motion can have only a
single equilibrium solution, corresponding to $f = 1/2$, since
fluctuations can always drive a transition from one local equilibrium to
another.  Thus, if we have a lattice with $l$ sites, the exact solution
of the dynamical equations for all CCF's of $3l$ or fewer particles
should only give a single solution.  It is interesting to consider the
effect that a finite lattice size would have on our discussion of the
2-particle BBGKY equations.  The only way in which a finite lattice size
would modify the equations is to change the coefficients $\phi_k (t)$ to
correspond to random walks on the finite lattice.  A particularly simple
example of this is the degenerate case where we have a lattice with only
a single vertex.  In this case, the outgoing particles from a collision
return immediately to the same vertex.  Thus, we have $\phi_k (1)= 1$
for all $k$, and of course $\phi_k (t)= 0$ for all $t > 1$.  This
modification of the coefficients has no effect on the exact equations at
a vertex, Eqs.~(\ref{eq:2equilibrium1}) and (\ref{eq:2equilibrium2}), so
$f = 1/2$ is still a solution of the equilibrium equations.  However,
using the modified values for $\phi$, the 2-particle BBGKY equation
(\ref{eq:final}) becomes
\[
\zeta (f) = -4 + 42 f-42 f^2.
\]
This equation has two solutions, which give spurious equilibria
analogous to those encountered previously on the infinite lattice.
Thus, although the finite size effects remove the extra physical
equilibria, which we only expect to exist in the thermodynamic limit,
these effects leave the spurious solutions of the BBGKY-truncated
equilibrium equations intact.  An interesting question, which we will
address in future work, is at precisely what lattice size the
thermodynamic equilibria first appear in the 2-particle BBGKY
approximation.  An answer to this and related questions might shed light
on the relationship between $i$-particle correlations and fluctuation
scales.

\section{Conclusions}

We have described an NSDB lattice gas model for Schl\"{o}gl's second
chemical reaction.  We derived a self-consistent set of equations for
its exact homogeneous equilibria, solved these equations in the
two-particle BBGKY approximation, and compared the results to numerical
experiment.  We found that this approximation describes the equilibria
far more accurately than the Boltzmann approximation, but we also noted
that it can give rise to spurious solutions to the equilibrium equations
which can only be removed by including effects due to three-particle
correlations.

The possibility of the existence of spurious solutions of the
two-particle BBGKY equations was raised by Bussemaker et.\
al.~\cite{bib:bed} The method they used to solve these equations was an
iterative approximation method which was not well suited to recognizing
the existence of multiple solutions.  The use in this paper of a
diagrammatic formalism to describe the time development of the
correlations made it possible to write the BBGKY-truncated equilibrium
equations in a closed form which was amenable to numerical solution.  It
would be interesting to extend the diagrammatic analysis described here
to higher-order truncations of the BBGKY hierarchy.

The physically meaningful solutions of these BBGKY-truncated equilibrium
equations provide an accurate description of the non-Gibbsian
equilibrium of this lattice gas.  The next step in this program of study
will be to expand about this non-Gibbsian equilibria in Knudsen number,
thereby generalizing the usual Chapman-Enskog analysis.  In this way,
the full reaction-diffusion equation, Eq.~(\ref{eq:sde}) will be
derived, including the renormalized diffusion coefficient.  This work is
in progress~\cite{bib:btnext}.

\section*{Acknowledgements}
One of us (BMB) would like to acknowledge helpful conversations with
Professor M.H. Ernst.  In addition, he would like to acknowledge the
hospitality of the Center for Computational Science at Boston
University, and the Information Mechanics Group at the M.I.T. Laboratory
for Computer Science.  This work was supported in part by the divisions
of Applied Mathematics of the U.S.  Department of Energy (DOE) under
contracts DE-FG02-88ER25065 and DE-FG02-88ER25066, and in part by the
U.S. Department of Energy (DOE) under cooperative agreement
DE-FC02-94ER40818.

\clearpage
\listoftables

\clearpage
\begin{table}
\begin{tabular}{|c|c||c|c|c|c|}
\hline
\multicolumn{2}{|c|}{$A(s\rightarrow s')$} & \multicolumn{4}{|c|}{$|s|$} \\
\cline{3-6}
\multicolumn{2}{|c|}{ } & $0$ & $1$ & $2$ & $3$ \\
\hline
\hline
  & $0$ & $\mixten{P}{0}{0}$
        & $\mixten{P}{0}{1}/3$
        & $\mixten{P}{0}{2}/3$
        & $\mixten{P}{0}{3}$ \\
\cline{2-6}
  & $1$ & $\mixten{P}{1}{0}$
        & $\mixten{P}{1}{1}/3$
        & $\mixten{P}{1}{2}/3$
        & $\mixten{P}{1}{3}$ \\
\cline{2-6}
$|s'|$
  & $2$ & $\mixten{P}{2}{0}$
        & $\mixten{P}{2}{1}/3$
        & $\mixten{P}{2}{2}/3$
        & $\mixten{P}{2}{3}$ \\
\cline{2-6}
  & $3$ & $\mixten{P}{3}{0}$
        & $\mixten{P}{3}{1}/3$
        & $\mixten{P}{3}{2}/3$
        & $\mixten{P}{3}{3}$ \\
\hline
\end{tabular}
\caption{Ensemble-averaged transition matrix}
\label{tab:eatm}
\end{table}

\begin{table}
\begin{tabular}{|c|c||c|c|c|c|}
\hline
\multicolumn{2}{|c|}{$\vop{\alpha}{\beta}$} & \multicolumn{4}{|c|}{$|\beta|$}\\
\cline{3-6}
\multicolumn{2}{|c|}{ } & $0$ & $1$ & $2$ & $3$ \\
\hline
\hline
  & $0$ & $1$
        & $0$
        & $0$
        & $0$ \\
\cline{2-6}
$|\alpha|$
  & $1$ & $0$
        & $1/3$
        & $0$
        & $0$ \\
\cline{2-6}
  & $2$ & $0$
        & $0$
        & $1/3$
        & $0$ \\
\cline{2-6}
  & $3$ & $0$
        & $0$
        & $0$
        & $1$ \\
\hline
\end{tabular}
\caption{Vertex coefficients for diffusive vertices}
\label{tab:vcdv}
\end{table}

\begin{table}
\begin{tabular}{|c|c||c|c|c|c|}
\hline
\multicolumn{2}{|c|}{$\vop{\alpha}{\beta}$} & \multicolumn{4}{|c|}{$|\beta|$}\\
\cline{3-6}
\multicolumn{2}{|c|}{ } & $0$ & $1$ & $2$ & $3$ \\
\hline
\hline
  & $0$ & $1$
        & $0$
        & $0$
        & $0$ \\
\cline{2-6}
$|\alpha|$
  & $1$ & $1/9$
        & $0$
        & $7/9$
        & $-14/9$ \\
\cline{2-6}
  & $2$ & $0$
        & $0$
        & $7/9$
        & $-14/9$ \\
\cline{2-6}
  & $3$ & $0$
        & $0$
        & $2/3$
        & $-4/3$ \\
\hline
\end{tabular}
\caption{Vertex coefficients for reactive vertices}
\label{tab:vcrv}
\end{table}

\clearpage
\listoffigures

\clearpage

\clearpage

\clearpage
\phantom{a}
\vspace{2.0truein}

\clearpage

\clearpage

\end{document}